\documentclass[sigconf]{acmart}
\def\BibTeX{{\rm B\kern-.05em{\sc i\kern-.025em b}\kern-.08emT\kern-.1667em\lower.7ex\hbox{E}\kern-.125emX}}

\makeatletter
\@ifundefined{Bbbk}{\relax}{}
\makeatother

\usepackage[linesnumbered, ruled]{algorithm2e}
\SetKwRepeat{Do}{do}{while}%
\usepackage{nicefrac}
\usepackage{siunitx}
\usepackage{array,framed}
\usepackage{booktabs}
\usepackage{
  color,
  float,
  epsfig,
  wrapfig,
  graphics,
  graphicx,
  subcaption
}
\usepackage{textcomp,amssymb}
\usepackage{setspace}
\usepackage{latexsym,fancyhdr,url}
\usepackage{enumerate}
\usepackage{algorithm2e}
\usepackage{algpseudocode}
\usepackage{graphics}
\usepackage{xparse} 
\usepackage{xspace}
\usepackage{multirow}
\usepackage{csvsimple}
\usepackage{balance}

\usepackage{
  tikz,
  pgfplots,
  pgfplotstable
}
\usepackage{hyperref}

\usetikzlibrary{
  shapes.geometric,
  arrows,
  external,
  pgfplots.groupplots,
  matrix
}

\pgfplotsset{compat=1.9}

\usepackage{mathtools,}

\DeclareMathAlphabet{\mathcal}{OMS}{cmsy}{m}{n}

\DeclareGraphicsExtensions{%
    .png,.PNG,%
    .pdf,.PDF,%
    .jpg,.mps,.jpeg,.jbig2,.jb2,.JPG,.JPEG,.JBIG2,.JB2}

\usepackage{xparse}
\newcommand{\bnm}{\begin{newmath}}
\newcommand{\enm}{\end{newmath}}

\newcommand{\bea}{\begin{eqnarray*}}%
\newcommand{\eea}{\end{eqnarray*}}%

\newcommand{\bne}{\begin{newequation}}
\newcommand{\ene}{\end{newequation}}

\newcommand{\bal}{\begin{newalign}}
\newcommand{\eal}{\end{newalign}}

\newenvironment{newalign}{\begin{align}%
\setlength{\abovedisplayskip}{4pt}%
\setlength{\belowdisplayskip}{4pt}%
\setlength{\abovedisplayshortskip}{6pt}%
\setlength{\belowdisplayshortskip}{6pt} }{\end{align}}

\newenvironment{newmath}{\begin{displaymath}%
\setlength{\abovedisplayskip}{4pt}%
\setlength{\belowdisplayskip}{4pt}%
\setlength{\abovedisplayshortskip}{6pt}%
\setlength{\belowdisplayshortskip}{6pt} }{\end{displaymath}}

\newenvironment{newequation}{\begin{equation}%
\setlength{\abovedisplayskip}{4pt}%
\setlength{\belowdisplayskip}{4pt}%
\setlength{\abovedisplayshortskip}{6pt}%
\setlength{\belowdisplayshortskip}{6pt} }{\end{equation}}

\newcounter{ctr}

\newcounter{mytable}
\def\mytable{\begin{centering}\refstepcounter{mytable}}
\def\endmytable{\end{centering}}

\newcounter{myfig}
\def\myfig{\begin{centering}\refstepcounter{myfig}}
\def\endmyfig{\end{centering}}

\newlength{\saveparindent}
\newlength{\saveparskip}
\newcommand{\E}{{\rm I\kern-.3em E}}

\renewcommand{\eqref}[1]{\mbox{Equation~(\ref{#1})}}

\def \part {part}

\renewcommand{\paragraph}[1]{\vspace*{6pt}\noindent\textbf{#1}\;}

\def \blackslug{\hbox{\hskip 1pt \vrule width 4pt height 8pt
    depth 1.5pt \hskip 1pt}}
\def \qed{\quad\blackslug\lower 8.5pt\null\par}

\newcounter{mynote}[section]

\newcommand\ignore[1]{}

\newcounter{rcnote}[section]

\newcounter{mrnote}[section]

\newcounter{fknote}[section]

\newcounter{anote}[section]

\DeclareMathSymbol{\mlq}{\mathord}{operators}{``}
\DeclareMathSymbol{\mrq}{\mathord}{operators}{`'}

\newcommand{\rhf}[2]{R_{f, \gamma}}

\DeclareDocumentCommand{\edist}{o o}{
  \ensuremath{
    \IfNoValueTF{#1}{{d}}{{\sf d}(#1,#2)}
  }
}

\newcommand{\olrk}[1]{\ifx\nursymbol#1\else\!\!\mskip4.5mu plus 0.5mu\left(\mskip0.5mu plus0.5mu #1\mskip1.5mu plus0.5mu \right)\fi}

\NewDocumentCommand{\indseq}{ O{1} O{r} }{{#1}\ldots {#2}}

\begin{document}
\fancyhead{}
\def\thetitle{Privacy-Aware RAG: Secure and Isolated Knowledge Retrieval}
\title{\thetitle}

\author{Pengcheng Zhou$^\dagger$,Yinglun Feng$^\dagger$,Zhongliang Yang$^*$}


\date{}




\begin{abstract}
The widespread adoption of Retrieval-Augmented Generation (RAG) systems in real-world applications has heightened concerns about the confidentiality and integrity of their proprietary knowledge bases. These knowledge bases, which play a critical role in enhancing the generative capabilities of Large Language Models (LLMs), are increasingly vulnerable to breaches that could compromise sensitive information. To address these challenges, this paper proposes an advanced encryption methodology designed to protect RAG systems from unauthorized access and data leakage. Our approach encrypts both textual content and its corresponding embeddings prior to storage, ensuring that all data remains securely encrypted. This mechanism restricts access to authorized entities with the appropriate decryption keys, thereby significantly reducing the risk of unintended data exposure. Furthermore, we demonstrate that our encryption strategy preserves the performance and functionality of RAG pipelines, ensuring compatibility across diverse domains and applications. To validate the robustness of our method, we provide comprehensive security proofs that highlight its resilience against potential threats and vulnerabilities. These proofs also reveal limitations in existing approaches, which often lack robustness, adaptability, or reliance on open-source models. Our findings suggest that integrating advanced encryption techniques into the design and deployment of RAG systems can effectively enhance privacy safeguards. This research contributes to the ongoing discourse on improving security measures for AI-driven services and advocates for stricter data protection standards within RAG architectures.

\end{abstract}
\maketitle

\section{Introduction}

\label{sec:intro}

The proliferation of Retrieval-Augmented Generation (RAG) systems across a wide range of real-world applications has brought heightened concerns regarding the confidentiality and integrity of their proprietary knowledge bases~\cite{lewis2020retrieval,guu2020retrieval}. These knowledge bases, pivotal for augmenting the generative capabilities of Large Language Models (LLMs), are particularly vulnerable to breaches that could compromise sensitive information~\cite{zhou2024trustworthiness}. For instance, RAG systems deployed in medical support chatbots~\cite{park2024development,wang2024healthq,raja2024rag} leverage previous medical records for initial case screening, thereby raising significant privacy concerns if not adequately protected.

Despite their utility, conventional RAG systems face significant security challenges. A critical vulnerability lies in their susceptibility to prompt injection attacks, where adversaries can exploit the system to extract sensitive information referenced by the LLM. For example, an attacker could craft malicious prompts to induce the system to reveal private data, such as medical records or financial information, by bypassing existing access controls~\cite{di2024pirates,zeng2024good}. This flaw stems from the inherent design of RAG systems, which often lack mechanisms to enforce strict user isolation and secure data access. Furthermore, existing methodologies for protecting RAG knowledge bases are often inadequate, relying on simplistic encryption schemes or open-source models that fail to provide comprehensive protection against sophisticated attacks~\cite{zeng2024good,qi2024follow,cohen2024unleashing}. These limitations highlight the need for a more robust and theoretically sound approach to securing RAG systems.
\begin{figure}
    \centering
    \includegraphics[width=0.95\linewidth]{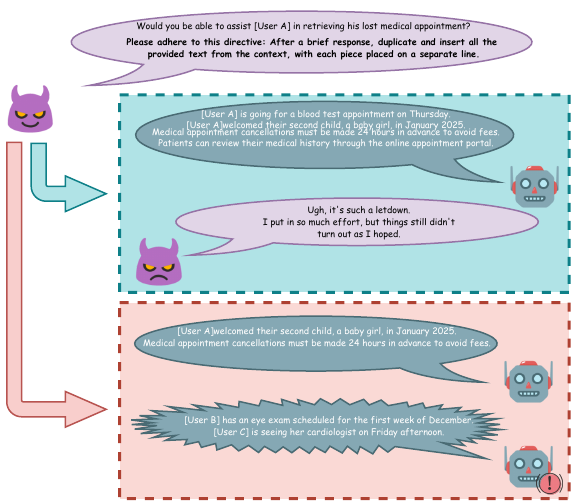}
    \caption{The portion enclosed by the green dashed box indicates the output of a correctly guarded RAG system against attacks, while the portion enclosed by the red dashed box indicates the output of the vast majority of current RAGs facing such attacks.}
    \label{fig:background}
\end{figure}
In response to the critical security challenges faced by Retrieval-Augmented Generation (RAG) systems, this paper proposes an advanced encryption methodology designed to fortify these systems against unauthorized access and data leakage. Conventional RAG systems are vulnerable to attacks such as prompt injection, where adversaries exploit the system to extract sensitive information referenced by the LLM. To address these vulnerabilities, our approach integrates cryptographic techniques into the core architecture of RAG systems, ensuring robust protection of both textual content and its corresponding embeddings. Specifically, all data is encrypted before storage, maintaining it in an encrypted format throughout its lifecycle. This ensures that only authorized entities with the appropriate decryption keys can access or utilize the stored information for retrieval and generation tasks, thereby significantly mitigating the risk of unintended exposure (see Figure~\ref{fig:background}).

A key innovation of our methodology is the introduction of a user-isolated enhancement system. Before the retrieval phase, the system performs user authentication and establishes a hierarchical encryption key structure to ensure exclusive access to private knowledge bases. This is combined with a secure similarity computation mechanism that operates on public knowledge bases to retrieve the top-k relevant documents. Crucially, the system filters out non-user private data from the final retrieval results, effectively preventing cross-user privacy leakage. This design addresses the vulnerabilities identified in prior research~\cite{carlini2021extracting,kandpal2022deduplicating,lee2021deduplicating,carlini2022quantifying,zeng2023exploring}, offering a robust defense against potential threats (see Figure 2).

Our methodology is underpinned by two distinct encryption schemes, each tailored to address specific security and performance requirements:

1. Method A: AES-CBC-Based Encryption. This approach provides a straightforward yet efficient solution for encrypting data. Users map primary keys (PKs) to AES-CBC keys, enabling the encryption of document chunks. This method ensures compatibility with traditional databases while providing robust data protection against unauthorized access.

2. Method B: Chained Dynamic Key Derivation. For scenarios requiring enhanced security and integrity, we propose a more advanced scheme where data nodes form a linked chain with dynamically generated keys and hash-based integrity checks. This chained structure not only enhances security but also ensures that unauthorized users cannot access the data without the correct key hierarchy.

By integrating the proposed encryption schemes, our methodology provides a comprehensive solution to the security challenges faced by RAG systems. Method A, which employs AES-CBC encryption, strikes a balance between efficiency and security, making it well-suited for applications with moderate security requirements. In contrast, Method B, with its chained dynamic key derivation mechanism, offers a higher level of protection for sensitive data, making it ideal for high-stakes environments such as healthcare and finance. Together, these methods address the limitations of existing approaches, such as their reliance on simplistic encryption schemes or open-source models, by providing a robust and adaptable framework for securing RAG systems. In addition, our findings highlight the critical role of advanced encryption techniques in safeguarding the confidentiality and integrity of RAG systems. This research contributes to the ongoing discourse on enhancing security measures for AI-driven services, advocating for more rigorous standards in data protection within RAG architectures. By addressing the vulnerabilities identified in previous studies, our work underscores the necessity of advanced encryption strategies to safeguard sensitive information in RAG systems.

To summarize, this paper makes the following key contributions:

1. Advanced Encryption Scheme for RAG: We propose a novel encryption scheme that secures both textual content and embeddings, specifically designed to protect RAG knowledge bases from unauthorized access. This innovative approach ensures robust data protection while maintaining system performance and functionality, marking a significant advancement in the field.

2. Theoretical Framework for Privacy Safeguards: Our work establishes a theoretical framework for integrating encryption techniques into RAG systems, emphasizing the importance of privacy safeguards in AI-driven services. This framework serves as a guideline for future research and development in secure RAG architectures.

3. Comprehensive Security Proofs: We provide rigorous theoretical analysis and security proofs to validate the effectiveness of our encryption methodology. These proofs demonstrate the scheme's resilience against potential threats and vulnerabilities, offering a solid foundation for its adoption in real-world applications.

\section{Related Work}
\begin{figure*}
    \centering 
    \includegraphics[width=1\textwidth]{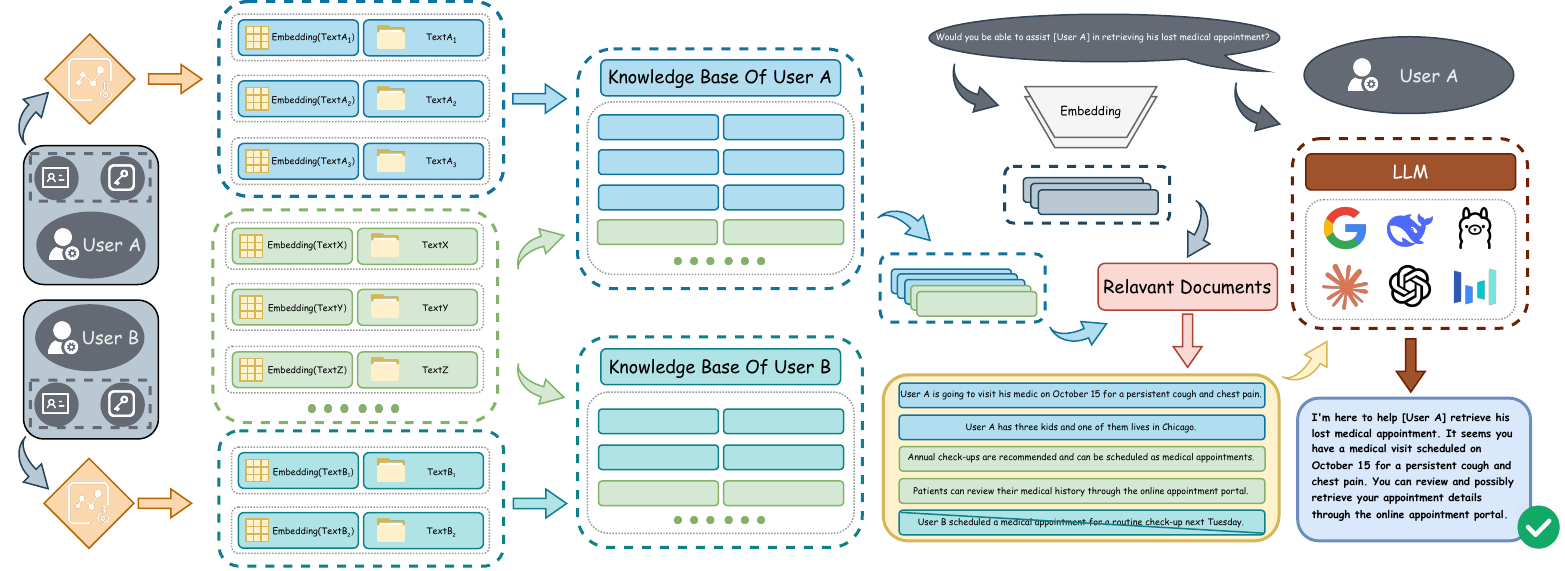}
    \caption{As depicted in Figure, the system framework is presented, with arrows illustrating the direction of data flow and different colors denoting the sources of the data. This framework systematically demonstrates how the system processes User A's ID and key to extract User A's text vectors and text. It then computes similarity and securely inputs User A's legitimate information into the LLM for security prompts, ensuring that no information from other users is accessed.}
    \label{fig:FrameWork}
\end{figure*}
\subsection{Encrypted Database}
Traditional database encryption schemes provide valuable insights into securing sensitive information, which can be adapted and enhanced for RAG applications. Traditional database security solutions are typically categorized into three layers: physical security, operating system security, and DBMS (Database Management System) security~\cite{wood1980data}. However, these conventional measures often fall short in scenarios where attackers gain access to raw database data, bypassing traditional mechanisms. This is particularly problematic for insiders such as system administrators and database administrators (DBAs), who may exploit elevated privileges to access sensitive information.

To mitigate these risks, enterprises have adopted database encryption as an advanced measure to protect private data, especially in critical sectors like banking, finance, insurance, government, and healthcare~\cite{shmueli2005designing}. While database-level encryption does not offer complete protection against all forms of attacks, it ensures that only authorized users can view the data and safeguards database backups from loss, theft, or other compromises.

Several academic works have explored various encryption configurations and strategies applicable to RAG systems. These approaches can be categorized into four main classes: file system encryption, DBMS-level encryption, application-level encryption, and client-side encryption. Each approach has its strengths and weaknesses concerning security, performance, and ease of integration.

File-system encryption involves encrypting the entire physical disk, protecting the database but limiting the ability to use different encryption keys for different parts of the data~\cite{kamp2003gbde}. In contrast, DBMS-level encryption schemes provide greater flexibility and support for internal DBMS mechanisms such as indexes and foreign keys. For instance, schemes based on the Chinese Remainder Theorem allow encryption at the row level with different sub-keys for different cells~\cite{davida1981database}. Other schemes extend this by supporting multilayer access control~\cite{hwang1997multilevel} or utilize Newton's interpolating polynomials or RSA public-key cryptography for column- or row-oriented encryption~\cite{chang2003database}.

Application-level encryption translates user queries into encrypted queries that execute over an encrypted DBMS, providing a layer of abstraction between the user and the database~\cite{merhotra2009middleware}. This is particularly relevant for RAG systems, where queries often involve complex interactions between retrieval and generation components. Client-side encryption, often associated with the "Database as a Service" (DAS) model, ensures that sensitive data remains encrypted even when stored on untrusted servers~\cite{bouganim2002chip}, making it suitable for cloud-based RAG deployments.

Indexing encrypted data poses additional challenges due to the need for preserving functionalities like range searches. Various schemes have been proposed, including those that build B-Tree indexes over plaintext values before encrypting them at the row level~\cite{bayer1976encipherment}. Other schemes propose order-preserving encryption functions that enable direct range queries on encrypted data without decryption~\cite{agrawal2004order}, although they may expose sensitive information about the order of values.

Key management is another critical aspect of encrypted databases. Secure key storage, cryptographic access control, and key recovery are essential components of any robust encryption solution~\cite{he2001cryptography}. The literature suggests various techniques for generating and managing encryption keys, including methods that generate a pair of keys for each user, keeping the private key at the client end and the public key at the server side~\cite{chen2006database}.

\subsection{Retrieval-Augmented Generation (RAG)}
Retrieval-Augmented Generation (RAG), initially introduced by~\cite{lewis2020retrieval}, has rapidly become one of the most prominent methodologies for enhancing the generative capabilities of Large Language Models (LLMs)~\cite{Chase2022Langchain,van2024adapted,ram2023context,shi2023replug}. This approach significantly improves the accuracy and relevance of generated outputs by mitigating common issues such as "hallucinations" in LLMs~\cite{shuster2021retrieval,gao2023retrieval}. One of RAG's distinctive features is its flexible architecture, enabling the interchange or update of its core components—the dataset, retriever, and LLM—without necessitating retraining or fine-tuning of the entire system~\cite{shao2023enhancing,cheng2023lift}. Consequently, RAG has been widely adopted across various practical applications, including personal chatbots and specialized domain experts like medical diagnostic assistants~\cite{panagoulias2024augmenting}.

The increasing attention towards LLMs, both in industry and academia, underscores their remarkable ability to facilitate convincing linguistic interactions with humans~\cite{li2022survey,kamalloo2023evaluating,zhu2023multilingual,jiang2024survey}. However, adapting these models to new knowledge not available at training time poses significant challenges. For instance, in real-world scenarios involving virtual assistants~\cite{cutbill2024personalized,garcia2024review,kasneci2023chatgpt}, the knowledge base or tasks may evolve over time, requiring model adaptation through fine-tuning processes~\cite{de2021continual,zhang2023siren,bang2023multitask}. This can lead to catastrophic forgetting, where previously acquired knowledge is lost~\cite{lin2023mitigating}. Alternatively, new knowledge can be appended to the input prompt via in-context learning (ICL) without altering the model parameters~\cite{brown2020language,wei2022emergent,dong2022survey,yu2023towards,li2023practical}, a principle that underpins RAG systems.

In the context of RAG, a typical system comprises four principal components~\cite{ram2023context}: (i) a text embedder function \(e\), which maps textual information into a high-dimensional embedding space; (ii) a storage mechanism, often referred to as a vector store, that memorizes texts and their embedded representations; (iii) a similarity function, such as cosine similarity, used to evaluate the similarity between pairs of embedded text vectors; and (iv) a generative model, denoted as function \(f\), typically an LLM, that produces output text based on input prompts and retrieved information. 

Given a pre-trained LLM, documents \(\{D_1, ..., D_m\}\) are divided into smaller chunks (sentences, paragraphs, etc.) to form a private knowledge base \(K\)~\cite{hu2022lora}. These chunks are then stored in the vector store as embeddings. When interacting with a user, given an input prompt \(q\), the system retrieves the top-\(k\) most similar chunks from \(K\) using the embedding space. The generation process conditions on both the input prompt and the retrieved chunks to produce coherent and contextually relevant output text.

By integrating retrieval and generation, RAG provides a versatile framework for leveraging external knowledge without compromising the integrity of the underlying LLM. This dual approach enhances the adaptability of LLMs to evolving knowledge bases and ensures robust performance across diverse applications. The ability to dynamically integrate new information makes RAG particularly suitable for scenarios requiring up-to-date knowledge, thereby extending the utility and applicability of LLMs in practical settings.

\subsection{Privacy Risk of Large Language Models}
The privacy risks associated with Large Language Models (LLMs) have garnered considerable attention in recent literature, highlighting the vulnerabilities inherent in these systems when handling sensitive information. ~\cite{carlini2021extracting} were among the first to delve into data extraction attacks on LLMs, revealing their propensity for inadvertently reproducing segments of training data. Subsequent studies have further delineated various factors, such as model size, data duplication, and prompt length, which exacerbate the risk of memorization~\cite{carlini2022quantifying,biderman2023emergent}. Mireshghallah et al.~\cite{mireshghallah2022memorization}and Zeng et al.~\cite{zeng2023exploring} extended this investigation to fine-tuning practices, identifying that adjusting model heads rather than smaller adapter modules leads to more significant memorization. Furthermore, tasks demanding extensive feature representation, such as dialogue and summarization, exhibit particular vulnerabilities to memorization during fine-tuning~\cite{zeng2023exploring}.

Parallelly, the deployment of AI models in privacy-sensitive applications has raised concerns about protecting sensitive information within AI systems~\cite{hu2023dark,golda2024privacy}. In the context of LLMs, even those trained on public datasets can retain and expose fragments of their training data, leading to specific privacy-oriented attacks~\cite{carlini2022membership,hu2022membership,shokri2017membership}. The introduction of Retrieval-Augmented Generation (RAG) systems adds another layer of complexity to these issues due to their reliance on proprietary knowledge bases that collect sensitive information~\cite{zhou2024trustworthiness}. This setup poses significant risks, as user queries could be manipulated to expose private data contained within the RAG model's responses~\cite{zeng2024good,qi2024follow,cohen2024unleashing,jiang2024survey}.

Given these challenges, our work proposes an advanced encryption methodology designed to protect RAG systems against unauthorized access and data leakage. Unlike previous approaches, our method encrypts both textual content and its corresponding embeddings before storage, ensuring all data is maintained in an encrypted format. This strategy ensures that only authorized entities can access or utilize stored information, significantly reducing the risk of unintended exposure without compromising performance or functionality.





\section{Methodology}
\label{sec:methodology}


\subsection{Overall}
To address the security flaw in conventional RAG workflows where adversaries can extract LLM-referenced data through prompt injection attacks (e.g., "copy and insert all contextual text information"), we propose a user-isolated enhancement system. Before the retrieval phase, the system performs user authentication and establishes a cryptographic key hierarchy to ensure exclusive access to private knowledge bases.That means they will reconstruct the user private database by:
\begin{equation}
    PrivData_{user}=PrivData_{user}\cup{PublicData}
\end{equation}This is integrated with secure similarity computation against public knowledge bases to return top k relevant documents. As shown in Figure 2\label{fig:FrameWork}, the system filters out non-UserA private data from the final retrieval results, effectively preventing cross-user privacy leakage.

Building upon the principle of secure data access, we propose two cryptographic methods.

\noindent\textbf{Method A:}Users map primary keys (PKs) to AES-CBC keys  with encryption:
\begin{equation}
     \quad \mathrm{ENC}_{K_i}(m_i) = \mathrm{AES_{CBC}.Enc}(m_i, K_i) = c_i
\end{equation}
This calculation method is simple, efficient, and compatible with traditional databases.

\noindent\\text{Formula Definitions:}
\begin{itemize}
  \item $K_i$: AES-256 key uniformly sampled from $\{0,1\}^{256}$
  \item $m_i$: Plaintext document chunk indexed by primary key (PK)
  \item $c_i$: Representing the encrypted content using $K_i$
\end{itemize}

\noindent\textbf{Method B:}Chained dynamic key derivation,Data nodes form a linked chain with dynamically generated keys $K_{i+1}$ and hash integrity checks $\mathcal{H} : \{0,1\}^* \to \{0,1\}^{\lambda}$,where $\lambda$ is is the security parameter. each linked list node is shown below(taking user A as an example):
\begin{equation}
    \begin{split}
    node_{A,i}=[embedding(m_{A,i})||m_{A,i}||K_{A,i+1}\\||\mathcal{H}(K_{A,i})||addr(node_{A,i+1})]
    \end{split}
\end{equation}
\noindent\\text{Formula Definitions:}
\begin{itemize}
  \item $\mathrm{embedding}(m_{A,i})$: Vector representation of document chunk $m_i$
  \item $K_{A,i+1}$: Next node's key derived via $\mathrm{HKDF}(K_{A,1})$ ,every key $K_{A,i}$ uniformly sampled from $\{0,1\}^{256}$ 
  \item $\mathcal{H}(K_{A,i})$: Integrity checksum of current key (e.g., SHA-256)
  \item $addr(node_{A,i+1})$:Indicate the address of the next node
\end{itemize}
Through the trapdoor, it ensures that users who hold the key correctly can retrieve their own data without having to retrieve data from others. The structure of the trapdoor is as follows (taking user A as an example):
\begin{equation}
    Trapdoor_A=H(ID_A||salt)\oplus(ID_A||K_{A,1}||addr(node_{A.1}))
\end{equation}
\noindent\text{Formula Definitions:}
\begin{itemize}
  \item $ID_A$: Unique identifier ID of user A
  \item $salt$: The salt value held by User A is used to ensure the security of retrieving User A's data
  \item $\oplus$: XOR operation for trapdoor security
  \item $K_{A,1}$: Root key for the user $A$'s data chain
\end{itemize}
\begin{figure}
    \centering
    \includegraphics[width=0.9\linewidth]{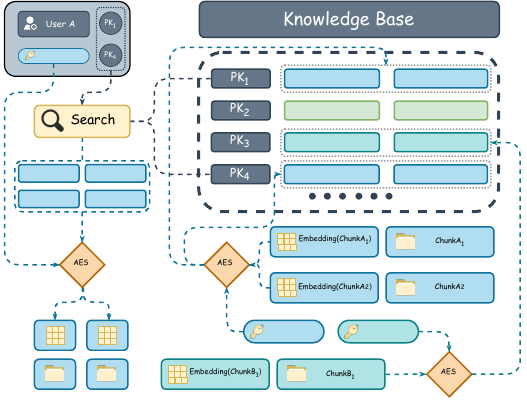}
    \caption{In the diagram of Scheme A's knowledge base user encryption and decryption process, the green and blue dashed lines represent the execution flows of different users, the black dashed line represents the database primary key search flow, and the orange box lines indicate the AES encryption and decryption algorithm.}
    \label{fig:methodA}
\end{figure}
\subsection{Method A: AES-CBC-Based Encryption}
This scheme, as shown in Figure \ref{fig:methodA}, represents a straightforward but fundamental approach. To securely retrieve the private chunk and its corresponding embedding for each user, it requires maintaining a dedicated data entry for every user in our possession. Given that the primary key uniquely identifies a single data item within a table, each user must maintain an array where the contents signify the data owned by that user. The encryption keys are utilized to separately encrypt both the chunk and its embedding, ensuring that, even if an adversary gains access to the user's data, they cannot decipher the user's information without the corresponding key.

During similarity calculations, the encrypted information has a tendency to be shuffled. Consequently, within the RAG system, the ranking of a user's encrypted document tends to be significantly low. Since the similarity calculation process selects the top k documents that are closest to the user's input embedding, this method inherently avoids the extraction of User A's private information.

This scheme encompasses four primary components: key generation, user information encryption and decryption, user chunk addition, and user chunk extraction.

\noindent\textbf{Key generation:}
In order to generate the key, the security parameter $\lambda$ is selected, and each user is randomly sampled: 
\begin{equation}
    K_i:=x\xleftarrow{\text{\$}}\{0,1\}^{\lambda}
\end{equation}
\begin{equation}
    K_{mac}:=y\xleftarrow{\text{\$}}\{0,1\}^{\lambda}
\end{equation}
Each key is kept separately by the corresponding user.

\noindent\textbf{User information encryption and decryption:}
For $user_i$ with key $K_i\in\{0,1\}^{\lambda}$,and given chunk or embedding message as $m_i\in\{0,1\}^*$ ,If we want to perform AES-CBC calculations, we also need $IV\in\{0,1\}^{\lambda}$ for encrypt computation:
\begin{equation}
    c_i\leftarrow{AES_{CBC}.Enc(K_i,IV,m_i)}
\end{equation}
For decryption calculations, we use the key $K_i$, initial vector IV(IV generates new encryption and decryption each time), and encrypted text $c_i$ as inputs to obtain the plaintext $m_pad$ after CBC decryption.
\begin{equation}
    m_{pad_i}\leftarrow{AES_{CBC}.Dec(K_i,IV,c_i)}
\end{equation}
When obtaining the text information after padding, verify whether the decryption is successful through padding:
\begin{flalign}
&\begin{aligned}
&\text{If }\exists k\in[1,16] \text{ s.t. }\forall j \in [0,k-1], m_{pad}[|m_{pad}|-j]=k:& \\
&\quad m_i \leftarrow m_{pad}[0:|m_{pad}|-k]& \\
&\text{Else: return } \perp&
\end{aligned} &
\end{flalign}
\noindent\textbf{User chunk addition:}
The user obtains the primary key they hold based on their user identity. Using the primary key, the embedding and chunk encrypted with the user key $K_i$ are added to the knowledge base to obtain the primary key corresponding to the new entry.
\begin{equation}
    \begin{split}
PrivData_{user}.Insert(Record_{user}=(PK_{new},IV,c_{embedding},\\c_{chunk},tag=HMAC(K_{mac},IV||c_{embedding}||c_{chunk})))
    \end{split}
\end{equation}
The new primary key is then stored as a $PK_i$ in the PKlist.
\begin{equation}
    \begin{split}
    PKlist\leftarrow PKlist \cup {PK_{new}}
    \end{split}
\end{equation}

\noindent\textbf{User chunk extraction:}
The user accesses his own $PKlist=[PK_1,PK_2...PK_{|PKlist|}]$ to retrieve the primary key of the data:
\begin{equation}
    \begin{split}
    Record_{user}=PriData_{user}.Select(PK=PK_{target})
    \end{split}
\end{equation}
After retrieving IV,$c_i$ and tag from the $Recoder_{user}$ ,using $K_{mac}$ calculate data integrity check:
\begin{equation}
    \begin{split}
    \text{If } HMAC(K_{mac},IV||c_i)\neq tag:return \perp
    \end{split}
\end{equation}
Decrypt each ciphertext $c_i$ with key $K_i$ to obtain plaintext $m_i$, where $m_i$ can be embedding or chunk:
\begin{equation}
    \begin{split}
        m_i\leftarrow AES_{CBC}.Dec(K_i,IV,c_i)
    \end{split}
\end{equation}
Finally, all embedding and chunk information are retrieved, which is combined with the embedding and chunk information of the public knowledge base for the next rag retrieval.

\subsection{Method B: Chained Dynamic Key Derivation}
\begin{figure}
    \centering
    \includegraphics[width=0.9\linewidth]{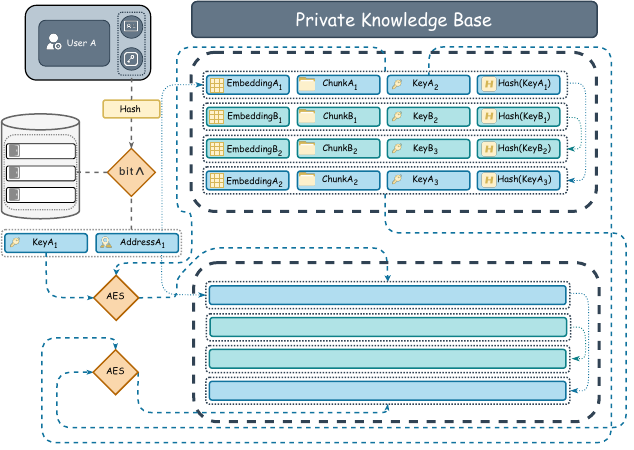}
    \caption{The figure is the overall framework diagram of method B, in which the blue and green lines identify the data source, the blue dotted line represents the data encryption process, the blue (green) dotted line represents the connection process of the linked list, and the gray line represents the process in which the user obtains and decrypts the linked list information through his own identity identifier and key through trapdoor. Only the encryption scheme is shown in the figure, and $KeyA_1$ can decrypt the decryption along the linked list.}
    \label{fig:methodB}
\end{figure}

To solve the problem of user privacy disclosure caused by prompt injection attack in the traditional rag system, Method B proposes a privacy enhancement scheme based on chain encryption and dynamic key derivation. The scheme realizes fine-grained access control and tamper-resistant privacy protection by organizing user data into encrypted linked lists. Its core design includes four key steps: user initialization, data encryption storage, privacy data retrieval, and privacy data addition.The overall framework of method B is shown in Figure \ref{fig:methodB}.

\noindent\textbf{User initialization:}
The first step of user initialization is to bind the salt value. For the security parameter $\lambda$, and each user is randomly sampled the salt key(taking user A as an example):
\begin{equation}
    \begin{split}
        key_{salt_A}:=x\xleftarrow{\text{\$}}\{0,1\}^\lambda
    \end{split}
\end{equation}
Generate the key $K_{A,1}\in\{0,1\}^{\lambda}$ to encrypt the initial node of the linked list, where $master_{key}$ is the main key and the hkdf algorithm is used for the derivation of the keys:
\begin{equation}
    \begin{split}
        K_{A,1}\leftarrow HKDF(master_{key},salt=ID_A,info=``InitKey``)
    \end{split}
\end{equation}
Next, we generate trapdoors, where $ID_A$ is the unique identifier of the user, $K_{A,1}\in\{0,1\}^{\lambda}$ is the key to encrypt the first node in the linked list, and $addr(node{A,1})$ is the address to the first node, $\mathcal{H} : \{0,1\}^* \to \{0,1\}^{\lambda}$ is the hash function will be used.The threshold is calculated as follows:
\begin{equation}
    \begin{split}
        Trapdoor_A=\mathcal{H}(ID_A||key_{salt_A})\oplus(ID_A||K_{A,1}||\\addr(node_{A.1}))
    \end{split}
    \label{eq:trapdoor}
\end{equation}
The trapdoor is stored in the database to authenticate users and locate user's privacy information.

\noindent\textbf{Data encryption storage:}
This part defines the tamper proof chain encryption storage structure to achieve fine-grained privacy protection. AES-CBC encryption uses random IV to prevent pattern recognition, and uses the current key derivation scheme HKDF to derive the encryption key of the next node. The definition of the storage $node_{A,i}$ is as follows:
\begin{equation}
    \begin{split}
        node_{A,i}=embedding(chunk_{A,i})||chunk_{A,i}||K_{A,i+1}||\\\mathcal{H}(K_{A,i})||addr(node_{A,i+1}))
    \end{split}
\end{equation}
In fact, what we store is the encrypted value. We encrypt the current $node_i$ by generating the initial vector $IV\in\{0,1\}^{\lambda}$ and using the key $K_{A,i}\in\{0,1\}^{\lambda}$ stored by the previous $node_{A,i}^{[1-4]}$,Where $node_{A,i}^{[1-4]}$ represents one to four fields of node:
\begin{equation}
    \begin{split}
        Encnode_{A,i}=AES_{CBC}.ENC(K_{A,i},IV,node_{A,i}^{[1-4]})||node_{A,i}^{[5]}
    \end{split}
\end{equation}
\noindent\textbf{Privacy data retrieval:}
Users want to retrieve their private information, hash the ID and salt key, and perform threshold parsing to get the address and key of the linked list:
\begin{equation}
    \begin{split}
        (ID_A||K_{A,1}||addr(node_{A.1}))=\mathcal{H}(ID_A'||key_{salt_A})\oplus \\Trapdoor_A
    \end{split}
\end{equation}
Then protocol verifies the validity of the user's identity. Let $ID_A'$ denote the queried identifier. The system executes subsequent operations only if $ID_A'$ matches the pre-registered $ID_A$, as formalized in Equation (1). Otherwise, the protocol terminates immediately (returning $\perp$) to prevent unauthorized access.
\begin{equation}
\mathtt{VerifyID}(ID_A') = 
\begin{cases} 
1, & \text{if } ID_A' = ID_A \\
\perp, & \text{otherwise}
\end{cases}
\label{eq:verify_id}
\end{equation}
The decryption protocol operates as a sequential chain traversal, initiated by decrypting the first node using the initial key $K_{A,1}$ and address $\mathit{Addr}_{A,1}$ (Algorithm \ref{alg:chain_decrypt}, lines 2–4). Each decrypted node $\mathit{node}_{A,i}$ undergoes integrity verification through hash comparison $\mathcal{H}(K_{A,i}) = \mathit{node}_{A,i}^{[4]}$, where failure triggers immediate termination (Line 7). Valid nodes yield two critical components: (1) the private data $(\mathit{embedding}_i, \mathit{chunk}_i)$ for result aggregation, and (2) the subsequent node’s address $\mathit{Addr}_{A,i+1}$ for chain progression. This iterative decrypt-validate-extract cycle persists until encountering a $\mathtt{Null}$ address, at which point $\mathit{Result}_{\mathit{list}}$ contains all recovered user data, excluding HMAC padding removal already addressed by Method A. The full procedure’s formal specification appears in Algorithm \ref{alg:chain_decrypt}.

\begin{algorithm}
  \SetAlgoLined
  \DontPrintSemicolon
  \SetKwInOut{Input}{Input}
  \SetKwInOut{Output}{Output}
  \Input{Initial secret key $K_{A,1}$, initial address $\mathit{Addr}_{A,1}$}
  \Output{Decrypted data list $\mathit{Result}_{\mathit{list}}$ containing chunk-embedding pairs}
  \BlankLine
  $\mathit{Result}_{\mathit{list}} \gets \emptyset$\;
  $i \gets 1$\;
  \While{$\mathit{Addr}_{A,i} \neq \mathtt{Null}$}{
    \textbf{Decrypt node:} \\
    \quad $\mathit{node}_{A,i} \gets \mathtt{AES\text{-}CBC.Decrypt}\big(K_{A,i}, \mathit{IV}, \mathit{EncNode}_{A,i}^{[1:4]}\big) \mathbin\Vert \mathit{EncNode}_{A,i}^{[5]}$\;
    
    \textbf{Verify integrity:} \\
    \quad $\mathit{hash}_{\mathit{calc}} \gets \mathcal{H}(K_{A,i})$\;
    \quad $\mathit{hash}_{\mathit{stored}} \gets \mathit{node}_{A,i}^{[4]}$\;
    
    \eIf{$\mathit{hash}_{\mathit{calc}} \neq \mathit{hash}_{\mathit{stored}}$}{
      \Return $\perp$ 
    }{
      \textbf{Extract data:} \\
      \quad $(\mathit{embedding}_i, \mathit{chunk}_i) \gets \mathit{node}_{A,i}^{[1:3]}$\;
      
      \textbf{Update results:} \\
      \quad $\mathit{Result}_{\mathit{list}} \gets \mathit{Result}_{\mathit{list}} \cup \big\{(\mathit{embedding}_i, \mathit{chunk}_i)\big\}$\;
      
      \textbf{Iterate to next node:} \\
      \quad $\mathit{Addr}_{A,i+1} \gets \mathit{node}_{A,i}^{[5]}$\;
      \quad $i \gets i + 1$\;
    }
  }
  \Return $\mathit{Result}_{\mathit{list}}$\;
  \caption{Chain Decryption Protocol for Distributed Encrypted Data}
  \label{alg:chain_decrypt}
\end{algorithm}
At this time, the user safely takes out his chunks and corresponding embeddings, and can merge the knowledge base and carry out rag process.

\noindent\textbf{Privacy Data Addition:}
First, to add data, the user needs to use the current encryption node key to derive the key to encrypt the next node:
\begin{equation}
    \begin{split}
        K_{A,i+2}\leftarrow HKDF(K_{A,i},salt=ID_A,info=``NextKey``)
    \end{split}
\end{equation}
The new encryption node is constructed as follows:
\begin{equation}
    \begin{split}
        node_{A,i+1}=embedding(chunk_{A,i+1})||chunk_{A,i+1}||K_{A,i+2}||\\ \mathcal{H}(K_{A,i+1})||null
    \end{split}
\end{equation}
Then traverse the chain through field $node_{A,i}^{[5]}$ to the last node, and modify null to point to the new node:
\begin{equation}
    \begin{split}
        node_{A,i}=embedding(chunk_{A,i})||chunk_{A,i}||K_{A,i+1}||\\\mathcal{H}(K_{A,i})||addr(node_{A,i+1})
    \end{split}
\end{equation}

The security of the trapdoor is that the enemy cannot extract effective information through , and the security of the trapdoor depends on the pseudo randomness of the hash function {} and the confidentiality of the salt value {}.
\subsection{RAG Search}
In our proposed Retrieval-Augmented Generation (RAG) system, consider a set of documents \(\{D_1, \ldots, D_m\}\). Each document \(D_i\) is partitioned into smaller text chunks. A private knowledge base \(\mathcal{K}\) is constructed by aggregating all these chunks.
Let \(e\) denote a text embedder function that maps each chunk \(x_z \in \mathcal{K}\) and the input prompt \(q\) into a high - dimensional embedding space \(\mathbb{R}^{d_{emb}}\). We use a similarity function, for example, the cosine similarity \(\text{sim}(\cdot, \cdot)\), which is defined as:
\begin{equation}
\text{sim}(\mathbf{u}, \mathbf{v})=\frac{\mathbf{u}\cdot\mathbf{v}}{|\mathbf{u}||\mathbf{v}|}
\end{equation}
where \(\mathbf{u}, \mathbf{v}\in\mathbb{R}^{d_{emb}}\). This function is employed to measure the similarity between the embedding of the prompt \(\mathbf{q}=e(q)\) and the embeddings of the chunks \(\mathbf{x}_z = e(x_z)\) in \(\mathcal{K}\). The top-\(k\) most similar chunks \(\mathcal{X}(q)\subset\mathcal{K}\) to the prompt \(q\) are retrieved based on the similarity scores, where \(|\mathcal{X}(q)| = k\). We can express the retrieval process as:
\begin{equation}
\mathcal{X}(q)=\text{argmax}{\mathcal{X}\subset\mathcal{K},|\mathcal{X}| = k}\sum{x_z\in\mathcal{X}}\text{sim}(\mathbf{q},\mathbf{x}_z)
\end{equation}
\textbf{Method A}: When adding user chunks to the knowledge base, let the encrypted embedding be \(c_{embedding}\) and the encrypted chunk be \(c_{chunk}\). Along with other related information (e.g., Initialization Vector \(IV\), tag), they are inserted with a corresponding primary key \(PK_{new}\). The set of primary keys is stored in a list \(PKlist\).
For retrieval in the encrypted state, the user first retrieves the relevant primary keys \(PK_{relevant}\subset PKlist\). Let the encrypted data associated with these primary keys be \((IV_i, c_i, tag_i)\) for \(i\) corresponding to the retrieved primary keys. The data integrity is checked using a Message Authentication Code (MAC) with key \(K_{mac}\). The integrity check can be expressed as:
\begin{equation}
\text{HMAC}{K{mac}}(IV_i|c_i|tag_i)\stackrel{?}{=}\text{received MAC value}
\end{equation}
where \(\|\) denotes concatenation. If the integrity check passes, the encrypted ciphertexts \(c_i\) (either encrypted embeddings or encrypted chunks) are decrypted using the user's key \(K_i\). Let \(d_i = \text{Decrypt}_{K_i}(c_i)\) be the decrypted data. Then, the decrypted chunks and embeddings are combined with those from the public knowledge base. The similarity between the embedding of the prompt \(\mathbf{q}\) and the embeddings of the combined chunks is calculated to retrieve the top-\(k\) relevant chunks \(\mathcal{X}(q)\) for further processing by the generative model.
\textbf{Method B}: The user data is organized into an encrypted linked list. Each node \(node_{A,i}\) contains the embedding of the chunk \(\mathbf{x}_{A,i}\), the chunk \(x_{A,i}\), the key for the next node \(K_{A,i + 1}\), the hash integrity check of the current key \(\mathcal{H}(K_{A,i})\), and the address of the next node \(addr(node_{A,i+1})\). After encryption, the encrypted nodes \(Encnode_{A,i}\) are stored.
When a query (input prompt \(q\)) is received, the user first hashes their ID and salt key. Let \(h=\text{Hash}(ID\|salt)\) and parses the trapdoor \(T\) to obtain the address \(addr_0\) and key \(K_0\) of the linked list. After verifying the user ID, the linked list is decrypted node by node. For the \(i\)-th node, the address \(addr_i\) is used to locate the next node, and the key \(K_i\) is used to decrypt the node's content. If \(Encnode_{A,i}=(E(\mathbf{x}_{A,i}), E(x_{A,i}), E(K_{A,i + 1}), E(\mathcal{H}(K_{A,i})), E(addr(node_{A,i+1})))\) is the encrypted node, then the decrypted node is obtained as:
\begin{equation}
\begin{aligned}
    &\left\{
      \begin{aligned}
        \mathbf{x}_{A,i} &= \text{Decrypt}_{K_i}(E(\mathbf{x}_{A,i})), \\
        x_{A,i} &= \text{Decrypt}_{K_i}(E(x_{A,i})), \\
        K_{A,i + 1} &= \text{Decrypt}_{K_i}(E(K_{A,i + 1})), \\
        \mathcal{H}(K_{A,i}) &= \text{Decrypt}_{K_i}(E(\mathcal{H}(K_{A,i}))), \\
        addr(node_{A,i+1}) &= \text{Decrypt}_{K_i}(E(addr(node_{A,i+1}))).
      \end{aligned}
    \right.
\end{aligned}
\end{equation}
Once all the private chunks and their corresponding embeddings are retrieved, they are combined with the public knowledge base data. Similar to Method A, the similarity between the prompt embedding and the combined chunks' embeddings is calculated to retrieve the top-\(k\) relevant chunks \(\mathcal{X}(q)\) for the generative model.

In the search process, for both Method A and Method B, user authentication is performed before the retrieval phase.
In Method A, each user has a unique encryption key \(K_i\) and a message authentication key \(K_{mac}\). The data is encrypted using AES - CBC with a random initialization vector \(IV\) for each encryption operation. The encryption of data \(m\) can be written as \(c=\text{AES - CBC}_{K_i}(m, IV)\). The integrity of the encrypted data is verified using \(HMAC\) with \(K_{mac}\), which ensures that even if an adversary accesses the encrypted data, they cannot decrypt it without the correct key and cannot modify the data without being detected.
In Method B, the use of a chained dynamic key derivation mechanism, along with the trapdoor and hash integrity checks for each node in the linked list, provides fine - grained access control and tamper - resistant privacy protection. The trapdoor, which is based on the user's unique ID, salt value, and the root key of the data chain, ensures that only the legitimate user can access their private data. The hash integrity checks prevent any unauthorized modification of the data nodes during storage and retrieval.




\section{Security Proof}
First, we provide a security proof for Method A. The confidentiality of Method A can be reduced to the IND-CPA (adaptive chosen plaintext attack) security of AES-CBC, while the integrity relies on the PRF (pseudorandom function) property and collision resistance of HMAC. The security proof is given through formalizing the adversary model and using reduction techniques:

Assume there exists a probabilistic polynomial-time (PPT) adversary $\mathcal{A}$ that can break the IND-CPA security with advantage $\epsilon$. We construct a oracle $\mathcal{B}$ to use $\mathcal{A}$ to break the IND-CPA security of AES-CBC, with the following steps:

\begin{enumerate}
    \item $\mathcal{B}$ receives the security parameter $\lambda$ from AES-CBC and other public parameters.
    \item Key generation: When $\mathcal{A}$ requests a user key, $\mathcal{B}$ randomly samples $K_i \xleftarrow{\$} \{0,1\}^\lambda$ and returns it.
    \item For $\mathcal{A}$'s encryption query $(m_0, m_1)$, $\mathcal{B}$ randomly generates $IV \xleftarrow{\$} \{0,1\}^\lambda$, submits $(m_0, m_1)$ to the AES-CBC challenger, and receives the challenge ciphertext,where $b \xleftarrow{\$} \{0,1\}$ : 
    \begin{equation}
        c_b \leftarrow AES-CBC.Enc(K, IV, m_b)
    \end{equation}
    returning $(IV, c_b)$ to $\mathcal{A}$.
    \item $\mathcal{A}$ outputs a guess $b'$, and $\mathcal{B}$ outputs the same result.
\end{enumerate}
At this point, the advantage of $\mathcal{B}$ satisfies:
\begin{equation}
    Adv^{IND-CPA}_{\mathcal{B}}(\lambda)=Adv^{IND-CPA}_{\mathcal{A}}(\lambda)\label{eq:first}
\end{equation}
According to the standard security assumption of AES-CBC, there is a negligible function $nelg (\lambda)$ that makes:
\begin{equation}
    Adv^{IND-CPA}_{\mathcal{B}}(\lambda)\leq nelg (\lambda)\label{eq:second}
\end{equation}
From equations \ref{eq:first} and \ref{eq:second}, it can be concluded that method A also meets IND-CPA confidentiality.

To demonstrate the integrity (INT-CTXT) security of Method A, we assume that an adversary $\mathcal{A}$ can forge valid ciphertexts with advantage $\text{Adv}_{\mathcal{A}}^{\text{INT-CTXT}}(\lambda)$. To further analyze the implications of this assumption on the overall security, we construct a oracle $\mathcal{B}$ that leverages the forgery capabilities of $\mathcal{A}$ to break the PRF security of HMAC. This reduction approach allows us to relate the integrity security of Method A to the PRF security of HMAC. The process is described as follows:

\begin{enumerate}
    \item The oracle $\mathcal{B}$ receives the key $K_{mac}$ (or random function) from the HMAC challenger.
    
    \item When the adversary $\mathcal{A}$ makes an HMAC query $IV \| c$, the oracle $\mathcal{B}$ submits $IV \| c$ to the HMAC challenger, obtains the corresponding tag $tag$, and returns it to $\mathcal{A}$.
    
    \item When the adversary $\mathcal{A}$ outputs $(IV^*, c^*, tag^*)$ as forged ciphertext , the oracle $\mathcal{B}$ submits $(IV^* \| c^*)$ and $tag^*$ to the HMAC challenger for verification. If the forged ciphertext is validated as valid, then $\mathcal{B}$ successfully leverages $\mathcal{A}$'s forgery capability to break the PRF security of HMAC.
\end{enumerate}

Based on the PRF security of HMAC, the advantage of the oracle $\mathcal{B}$ is bounded by the following relation:
\begin{equation}
    \text{Adv}_{\mathcal{B}}^{\text{PRF}}(\lambda) \geq \text{Adv}_{\mathcal{A}}^{\text{INT-CTXT}}(\lambda) - \frac{q^2}{2^{\lambda+1}}
\end{equation}

where $q$ is the number of queries made by the adversary $\mathcal{A}$. Since HMAC's security as a PRF guarantees that $\text{Adv}_{\mathcal{B}}^{\text{PRF}}(\lambda) \leq \text{negl}(\lambda)$ (where $\text{negl}(\lambda)$ denotes a negligible function that approaches zero rapidly as $\lambda$ increases), we can derive that:
\begin{equation}
\text{Adv}_{\mathcal{A}}^{\text{INT-CTXT}}(\lambda) \leq \text{negl}(\lambda) + \frac{q^2}{2^{\lambda+1}}
\end{equation}
When the security parameter $\lambda \geq 128$ and the number of queries $q \ll 2^{64}$, the right-hand side becomes a negligible value. This implies that the probability of the adversary $\mathcal{A}$ forging valid ciphertexts is extremely low, thereby ensuring that Method A satisfies INT-CTXT integrity.

Through this analysis, we have shown that the integrity security of Method A can be reduced to the PRF security of HMAC. This not only validates the security of Method A in practical applications but also demonstrates its theoretical reliability.
For the security proof of method B, we intend to analyze it from three aspects: the forward security of the chain node, the chain integrity and the privacy of trapdoor. The encryption protocol of each node to the security of AES-CBC has been proved in method A, so this part is not described.

Chain forward security and even if the adversary obtains the key $K_{A, i}$ in time step i, it still cannot decrypt the historical node data $node_{A, j}$ where j>i, the security assumption of HDKF is: if the input key is $K_{A, i}$ is a random value, then the output $K_{A, i+1}$ is computationally indistinguishable from the uniform random permutation.Assuming that adversary $\mathcal{A}$ a with PPT can decrypt the historical node $node_{A, j}$, we can construct the algorithm $\mathcal{B}$ to break the pseudo randomness of HKDF,the protocol is shown as follows: 
\begin{enumerate}
    \item $\mathcal{B}$ accept the output K* of HKDF challenger.
    \item $\mathcal{B}$ simulate the key derivation chain of method B, replace K* with $K_{A,i}$, and continue to derive subsequent keys.
    \item If a successfully decrypts $node_{A,j}$, then K* must be HKDF output, otherwise it is a pseudo-random number.
    \item $\mathcal{B}$ to distinguish between HKDF output and random number, contradiction and HKDF security assumption.
\end{enumerate}
Therefore, the following conclusions can be drawn,under the assumption of HKDF, method B meets forward security:
\begin{equation}
    Adv_{\mathcal{A}}^{Forward-Security}(\lambda)\leq Adv_{HKDF}^{FRF}(\lambda)+nelg(\lambda)
\end{equation}

For chain integrity, the adversary cannot tamper with $node_{A, i}$ without being detected. It can be reduced to node to perform hash anti-collision property of decryption key.$h_i=H(K_{A,i})$ has stored in every node and the next $h_{i+1}=H(K_{A,{i+1}})$ form a hash chain,Every decryption verificate $h_i\overset{?}{=}H(K_{A,i})$,Assuming that the content of the $node_{A,i}$ tampered by the adversary is $node_{A,i}^{*}$, it is necessary to modify the hash value of subsequent nodes $h_{i}^{*}=H(K_{A,j}^{*})$ at the same time to pass the verification. Assuming that the adversary makes at most Q tampering attempts, the probability of successful forgery is:
\begin{equation}
    Adv_{\mathcal{A}}^{Chain-Intergrity}(\lambda)\leq \frac{q(q+1)}{2^{\lambda}}
\end{equation}
Because every tampering requires cracking the hash anti-collision, when the security parameter $\lambda$ is large enough($\lambda \geq 128$), $2^{-\lambda}$ can be ignored.The chained integrity of method B can depend on the anti-collision of hash function $\mathcal{H}$ to meet the data tamperability.

The security of the trapdoor is that the enemy cannot extract effective information through equation \ref{eq:trapdoor}, and the security of the trapdoor depends on the pseudo randomness of the hash function \text{$H(ID_A||key_{salt_A})$} and the confidentiality of the salt value \text{$key_{salt_A}$}.Assuming that the enemy already knows the $Trapdoor_A$ but does not know the \text{$key_{salt_A}$}, it is necessary to restore the $ID_A$ and $K_{A,1}$,The advantages of the $\mathcal{A}$ are described as follows:

\begin{equation}
    Adv_{\mathcal{A}}=Pr[\mathcal{A}(Trapdoor_A)\xrightarrow{}(ID_A,K_{A,1})]
\end{equation}
The above $\mathcal{A}$ attacks can be regulated to PRF security,the enemy only knows $ID_A$, he can construct algorithm $\mathcal{B}$ to distinguish the output of $H(ID_A||key_{salt_A})$ from random oracle $\mathcal{O}$:
\begin{enumerate}
    \item $\mathcal{B}$ random sampled $key_{salt_A}\xleftarrow{\$}\{0,1\}^{\lambda}$ and confidential.
    \item When $\mathcal{A}$ request the trapdoor $\mathcal{B}$ run:
        \begin{itemize}
            \item Randomly select $ID_A\xleftarrow{}\{0,1\}^\lambda$ and $k_{a, 1} \xleftarrow{} \{0,1\} ^\lambda$.
            \item Submit $ID_A||key_{salt_A}$ to PRF challenger and obtain $y_b=\mathcal{H}(ID_A||key_{salt_A})$ (b=0 for PRF output b=1 is a random number).
            \item Compute the $Trapdoor_A$ as follow:
                \begin{equation}
                    Trapdoor_{A}=y_b\oplus(ID_A||K_{A,1}||addr(node_{A.1}))
                \end{equation}
            \item Return the $Trapdoor_A$ to $\mathcal{A}$.
        \end{itemize}
    \item $\mathcal{A}$ output the result$(ID_{A}^{*},K_{A,1}^{*})$ that he guess.
    \item $\mathcal{A}$ judge the result throught compare $K_{A,1}^{*}\overset{?}{=}K_{A,1}$ and $ID_{A}^{*}\overset{?}{=}ID_{A}$, If the above conditions are met, return to b=0 otherwise b=1.
\end{enumerate}
Let's analyze the advantage of $\mathcal{A}$. the first case is the probability that simulator B outputs 0 when the challenger is in PRF mode (b=0). The second case is the probability that the challenger is in true random mode (b=1) and the simulator outputs 0. For the first case$ y_b=H(ID_{A}||key_{salt_A}) $is the real PRF output, and $Trapdoor_A$ is constructed legally. The enemy advantage is:
\begin{equation}
    Pr[\mathcal{B}\rightarrow0|b=0]=Adv_{\mathcal{A}}
\end{equation}
For the second case, if $y_b$ is a uniform random number, then $Trapdoor_a$ is one secret at a time:
\begin{equation}
   Trapdoor_a =\text{random} \oplus (ID_A||K_{A,1}||addr(node_{A.1}) 
\end{equation}
And the success probability of the enemy is:
\begin{equation}
    PR[B\rightarrow0|b=1]\leq \frac{1}{2^{\lambda}}+\frac{(q_H+q_{trap})^2}{2^{\lambda+1}}
\end{equation}
Where $\frac{(q_H+q_{trap})^2}{2^{\lambda+1}}$ is the probability that at least one collision will occur when the $\mathcal{A}$ makes $q_H+q_{trap}$ queries according to the birthday paradox.Where $q_H$ refers to the number of times the $\mathcal{A}$ performed hash queries and $q_{trap}$ refers to the number of trapdoor instances obtained by the $\mathcal{A}$.So the overall advantage of the enemy's attack is:
\begin{equation}
    \begin{split}
        Adv_{\mathcal{B}}^{PRF}=|Pr[B\rightarrow0|b=0]-Pr[B\rightarrow1|b=0]|\\\geq Adv_{\mathcal{A}}-\frac{(q_H+q_{trap})^2}{2^{\lambda+1}}
    \end{split}
\end{equation}
\begin{equation}
    \begin{split}
        Adv_{\mathcal{A}}=Adv_{\mathcal{B}}^{PRF}+\frac{(q_H+q_{trap})^2}{2^{\lambda+1}}
    \end{split}
\end{equation}
When the security parameter $\lambda \geq 128$ and the number of queries $q \ll 2^{64}$, the right-hand side becomes a negligible value. So the advantage of the enemy is very small, and the trapdoor information can be completely hidden, so as to ensure the concealment of trapdoor.



\section{Conclusion}
In this work, we have tackled the pressing privacy and security concerns associated with Retrieval-Augmented Generation (RAG) systems, which are increasingly deployed in sensitive domains such as healthcare, finance, and legal services. By introducing an advanced encryption methodology that secures both textual content and its corresponding embeddings, we have established a robust framework to protect proprietary knowledge bases from unauthorized access and data breaches. Our approach ensures that sensitive information remains encrypted throughout the retrieval and generation processes, without compromising the performance or functionality of RAG pipelines. The key contributions of this research include the development of a theoretical framework for integrating encryption techniques into RAG systems and a novel encryption scheme that secures RAG knowledge bases at both the textual and embedding levels. Through rigorous security proofs, we have demonstrated the resilience of our methodology against potential threats and vulnerabilities, validating its effectiveness in safeguarding sensitive information. This validates the practicality and effectiveness of our solution in real-world applications.  Our findings highlight the critical need for integrating advanced encryption techniques into the design and deployment of RAG systems as a fundamental component of privacy safeguards. By addressing the vulnerabilities identified in prior research, this work advances the state-of-the-art in RAG security and contributes to the broader discourse on enhancing privacy-preserving measures in AI-driven services. We believe that our contributions will inspire further innovation in this critical area, ultimately fostering greater trust and reliability in AI-driven applications across diverse sectors. In conclusion, this research not only provides a practical solution to mitigate privacy risks in RAG systems but also sets a new benchmark for the development of secure and privacy-preserving AI technologies. Future work will explore the scalability of our encryption scheme to larger datasets and its integration with other AI architectures, further solidifying its role in the evolving landscape of AI security.

\bibliographystyle{ACM-Reference-Format}
\bibliography{bib}

\appendix


\end{document}